\documentstyle[12pt]{article}
\newcommand{\nc}{\newcommand}
\nc{\renc}{\renewcommand}

\nc{\half}{{\textstyle{1\over2}}}
\nc{\etal}{\mbox{\it et al. }}
\nc{\ie}{{\it i.e.}}
\nc{\eg}{{\it e.g.}}

\renc{\thefootnote}{\arabic{footnote}}
\nc{\capt}[1]{{\bf Figure.} {\small\sl #1}}

\nc{\eqs}[2]{\mbox{Eqs.~(\ref{#1},\,\ref{#2})}}
\nc{\eq}[1]{\mbox{Eq.~(\ref{#1})}}

\nc{\figs}[2]{\mbox{Figs.~(\ref{#1},\,\ref{#2})}}
\nc{\fig}[1]{\mbox{Fig~.(\ref{#1})}}

\nc{\tag}[1]{\label{#1} \marginpar{{\footnotesize #1}}}
\nc{\mtag}[1]{\label{#1} \mbox{\marginpar{{\footnotesize #1}}}}
\renc{\baselinestretch}{1.5}
\newlength{\overeqskip}
\newlength{\undereqskip}
\nc{\be}[1]{\begin{equation} \mbox{$\label{#1}$}}
\nc{\bea}[1]{\begin{eqnarray} \mbox{$\label{#1}$}}
\nc{\Section}[2]{\section{#2}\label{#1}}
\nc{\Bibitem}[1]{\bibitem{#1}}
\nc{\Label}[1]{\label{#1}}

\nc{\eea}{\vspace{\undereqskip}\end{eqnarray}}
\nc{\ee}{\vspace{\undereqskip}\end{equation}}
\nc{\bdm}{\begin{displaymath}}
\nc{\edm}{\end{displaymath}}
\nc{\dpsty}{\displaystyle}
\nc{\bc}{\begin{center}}
\nc{\ec}{\end{center}}
\nc{\ba}{\begin{array}}
\nc{\ea}{\end{array}}
\nc{\bab}{\begin{abstract}}
\nc{\eab}{\end{abstract}}
\nc{\btab}{\begin{tabular}}
\nc{\etab}{\end{tabular}}
\nc{\bit}{\begin{itemize}}
\nc{\eit}{\end{itemize}}
\nc{\ben}{\begin{enumerate}}
\nc{\een}{\end{enumerate}}
\nc{\bfig}{\begin{figure}}
\nc{\efig}{\end{figure}}
\nc{\arreq}{&\!=\!&}
\nc{\arrmi}{&\!-\!&}
\nc{\arrpl}{&\!+\!&}
\nc{\arrap}{&\!\!\!\approx\!\!\!&}
\nc{\non}{\nonumber\\*}
\nc{\align}{\!\!\!\!\!\!\!\!&&}

\def\lsim{\; \raise0.3ex\hbox{$<$\kern-0.75em
      \raise-1.1ex\hbox{$\sim$}}\; }
\def\gsim{\; \raise0.3ex\hbox{$>$\kern-0.75em
      \raise-1.1ex\hbox{$\sim$}}\; }
\nc{\DOT}{\hspace{-0.08in}{\bf .}\hspace{0.1in}}
\nc{\Laada}{\hbox {$\sqcap$ \kern -1em $\sqcup$}}
\nc\loota{{\scriptstyle\sqcap\kern-0.55em\hbox{$\scriptstyle\sqcup$}}}
\nc\Loota{{\sqcap\kern-0.65em\hbox{$\sqcup$}}}
\nc\laada{\Loota}
\nc{\qed}{\hskip 3em \hbox{\BOX} \vskip 2ex}

\nc{\real}{{\rm I \! R}}
\nc{\Z}{{\sf Z \!\!\! Z}}
\nc{\complex}{{\rm C\!\!\! {\sf I}\,\,}}
\def\bigid{\leavevmode\hbox{\small1\kern-3.8pt\normalsize1}}
\def\id{\leavevmode\hbox{\small1\kern-3.3pt\normalsize1}}
\nc{\slask}{\!\!\!/}
\nc{\bis}{{\prime\prime}}
\nc{\pa}{\partial}
\nc{\na}{\nabla}
\nc{\ra}{\rangle}
\nc{\la}{\langle}
\nc{\goto}{\rightarrow}
\nc{\swap}{\leftrightarrow}

\nc{\EE}[1]{ \mbox{$\cdot10^{#1}$} }
\nc{\abs}[1]{\left|#1\right|}
\nc{\at}[2]{\left.#1\right|_{#2}}
\nc{\norm}[1]{\|#1\|}
\nc{\abscut}[2]{\Abs{#1}_{\scriptscriptstyle#2}}
\nc{\vek}[1]{{\rm\bf #1}}
\nc{\integral}[2]{\int\limits_{#1}^{#2}}
\nc{\inv}[1]{\frac{1}{#1}}
\nc{\dd}[2]{{{\partial #1}\over{\partial #2}}}
\nc{\ddd}[2]{{{{\partial}^2 #1}\over{\partial {#2}^2}}}
\nc{\dddd}[3]{{{{\partial}^2 #1}\over
        {\partial #2 \partial #3}}}
\nc{\dder}[2]{{{d #1}\over{d #2}}}
\nc{\ddder}[2]{{{d^2 #1}\over{d {#2}^2}}}
\nc{\dddder}[3]{{d^2 #1}\over
        {d #2 d #3}}
\nc{\dx}[1]{d\,^{#1}x}
\nc{\dy}[1]{d\,^{#1}y}
\nc{\dz}[1]{d\,^{#1}z}
\nc{\dl}[1]{\frac{d\,^{#1}l}{(2\pi)^{#1}}}
\nc{\dk}[1]{\frac{d\,^{#1}k}{(2\pi)^{#1}}}
\nc{\dq}[1]{\frac{d\,^{#1}q}{(2\pi)^{#1}}}

\nc{\cc}{\mbox{$c.c.$ }}
\nc{\hc}{\mbox{$h.c.$ }}
\nc{\cf}{cf.\ }
\nc{\erfc}{{\rm erfc}}
\nc{\Tr}{{\rm Tr\,}}
\nc{\tr}{{\rm tr\,}}
\nc{\pol}{{\rm pol}}
\nc{\sign}{{\rm sign}}
\nc{\bfT}{{\bf T }}

\def\GeV{{\rm\ GeV}}

\nc{\cA}{{\cal A}}
\nc{\cB}{{\cal B}}
\nc{\cD}{{\cal D}}
\nc{\cE}{{\cal E}}
\nc{\cG}{{\cal G}}
\nc{\cH}{{\cal H}}
\nc{\cL}{{\cal L}}
\nc{\cO}{{\cal O}}
\nc{\cT}{{\cal T}}
\nc{\cN}{{\cal N}}
\nc{\rvac}[1]{|{\cal O}#1\rangle}
\nc{\lvac}[1]{\langle{\cal O}#1|}
\nc{\rvacb}[1]{|{\cal O}_\beta #1\rangle}
\nc{\lvacb}[1]{\langle{\cal O}_\beta #1 |}
\nc{\bb}{\bar{\beta}}
\nc{\bt}{\tilde{\beta}}
\nc{\ctH}{\tilde{\cal H}}
\nc{\chH}{\hat{\cal H}}
\nc{\1}{\aa}
\nc{\2}{\"{a}}
\nc{\3}{\"{o}}
\nc{\4}{\AA}
\nc{\5}{\"{A}}
\nc{\6}{\"{O}}
\nc{\al}{\alpha}
\nc{\g}{\gamma}
\nc{\Del}{\Delta}
\nc{\e}{\epsilon}
\nc{\eps}{\epsilon}
\nc{\lam}{\lambda}
\nc{\om}{\omega}
\nc{\Om}{\Omega}
\nc{\ve}{\varepsilon}
\nc{\mn}{{\mu\nu}}
\nc{\k}{\kappa}
\nc{\vp}{\varphi}

\nc{\advp}[3]{{\it  Adv.\ in\ Phys.\ }{{\bf #1} {(#2)} {#3}}}
\nc{\annp}[3]{{\it  Ann.\ Phys.\ (N.Y.)\ }{{\bf #1} {(#2)} {#3}}}
\nc{\apl}[3]{{\it  Appl. Phys. Lett. }{{\bf #1} {(#2)} {#3}}}
\nc{\apj}[3]{{\it  Ap.\ J.\ }{{\bf #1} {(#2)} {#3}}}
\nc{\apjl}[3]{{\it  Ap.\ J.\ Lett.\ }{{\bf #1} {(#2)} {#3}}}
\nc{\app}[3]{{\it Astropart.\ Phys.\ }{{\bf #1} {(#2)} {#3}}}
\nc{\cmp}[3]{{\it  Comm.\ Math.\ Phys.\ }{{ \bf #1} {(#2)} {#3}}}
\nc{\cqg}[3]{{\it  Class.\ Quant.\ Grav.\ }{{\bf #1} {(#2)} {#3}}}
\nc{\epl}[3]{{\it  Europhys.\ Lett.\ }{{\bf #1} {(#2)} {#3}}}
\nc{\ijmp}[3]{{\it Int.\ J.\ Mod.\ Phys.\ }{{\bf #1} {(#2)} {#3}}}
\nc{\ijtp}[3]{{\it Int.\ J.\ Theor.\ Phys.\ }{{\bf #1} {(#2)} {#3}}}
\nc{\jmp}[3]{{\it  J.\ Math.\ Phys.\ }{{ \bf #1} {(#2)} {#3}}}
\nc{\jpa}[3]{{\it  J.\ Phys.\ A\ }{{\bf #1} {(#2)} {#3}}}
\nc{\jpc}[3]{{\it  J.\ Phys.\ C\ }{{\bf #1} {(#2)} {#3}}}
\nc{\jap}[3]{{\it J.\ Appl.\ Phys.\ }{{\bf #1} {(#2)} {#3}}}
\nc{\jpsj}[3]{{\it J.\ Phys.\ Soc.\ Japan\ }{{\bf #1} {(#2)} {#3}}}
\nc{\lmp}[3]{{\it Lett.\ Math.\ Phys.\ }{{\bf #1} {(#2)} {#3}}}
\nc{\mpl}[3]{{\it  Mod.\ Phys.\ Lett.\ }{{\bf #1} {(#2)} {#3}}}
\nc{\ncim}[3]{{\it  Nuov.\ Cim.\ }{{\bf #1} {(#2)} {#3}}}
\nc{\np}[3]{{\it  Nucl.\ Phys.\ }{{\bf #1} {(#2)} {#3}}}
\nc{\npps}[3]{{\it  Nucl.\ Phys.\ Proc.\ Suppl.\ }{{\bf #1} {(#2)} {#3}}}
\nc{\pr}[3]{{\it Phys.\ Rev.\ }{{\bf #1} {(#2)} {#3}}}
\nc{\pra}[3]{{\it  Phys.\ Rev.\ A\ }{{\bf #1} {(#2)} {#3}}}
\nc{\prb}[3]{{\it  Phys.\ Rev.\ B\ }{{{\bf #1} {(#2)} {#3}}}}
\nc{\prc}[3]{{\it  Phys.\ Rev.\ C\ }{{\bf #1} {(#2)} {#3}}}
\nc{\prd}[3]{{\it  Phys.\ Rev.\ D\ }{{\bf #1} {(#2)} {#3}}}
\nc{\prl}[3]{{\it Phys.\ Rev.\ Lett.\ }{{\bf #1} {(#2)} {#3}}}
\nc{\pl}[3]{{\it  Phys.\ Lett.\ }{{\bf #1} {(#2)} {#3}}}
\nc{\prep}[3]{{\it Phys.\ Rep.\ }{{\bf #1} {(#2)} {#3}}}
\nc{\prsl}[3]{{\it Proc.\ R.\ Soc.\ London\ }{{\bf #1} {(#2)} {#3}}}
\nc{\ptp}[3]{{\it  Prog.\ Theor.\ Phys.\ }{{\bf #1} {(#2)} {#3}}}
\nc{\ptps}[3]{{\it  Prog\ Theor.\ Phys.\ suppl.\ }{{\bf #1} {(#2)} {#3}}}
\nc{\physa}[3]{{\it  Physica\ A\ }{{\bf #1} {(#2)} {#3}}}
\nc{\physb}[3]{{\it  Physica\ B\ }{{\bf #1} {(#2)} {#3}}}
\nc{\phys}[3]{{\it Physica\ }{{\bf #1} {(#2)} {#3}}}
\nc{\rmp}[3]{{\it  Rev.\ Mod.\ Phys.\ }{{\bf #1} {(#2)} {#3}}}
\nc{\rpp}[3]{{\it Rep.\ Prog.\ Phys.\ }{{\bf #1} {(#2)} {#3}}}
\nc{\sjnp}[3]{{\it Sov.\ J.\ Nucl.\ Phys.\ }{{\bf #1} {(#2)} {#3}}}
\nc{\spjetp}[3]{{\it Sov.\ Phys.\ JETP\ }{{\bf #1} {(#2)} {#3}}}
\nc{\yf}[3]{{\it Yad.\ Fiz.\ }{{\bf #1} {(#2)} {#3}}}
\nc{\zetp}[3]{{\it Zh.\ Eksp.\ Teor.\ Fiz.\  }{{\bf #1}  {(#2)} {#3}}}
\nc{\zp}[3]{{\it Z.\ Phys.\ }{{\bf #1} {(#2)} {#3}}}
\nc{\ibid}[3]{{\sl ibid.\ }{{\bf #1} {#2} {#3}}}
\nc{\rf}[1]{(\ref{#1})}
\nc{\nn}{\nonumber \\*}
\nc{\bfB}{\bf{B}}
\nc{\bfv}{\bf{v}}
\nc{\bfx}{\bf{x}}
\nc{\bfy}{\bf{y}}
\nc{\vx}{\vec{x}}
\nc{\vy}{\vec{y}}
\nc{\oB}{\overline{B}}
\nc{\oI}{\overline{I}}
\nc{\oR}{\overline{R}}
\nc{\rar}{\rightarrow}
\nc{\ti}{\times}
\nc{\slsh}{\hskip-5pt/}
\nc{\sm}{Standard~Model~}
\nc{\MP}{M_{\rm Pl}}
\nc{\tp}{t_{\rm Pl}}
\nc{\ave}{\bar{E}}

\nc{\eff}{{\rm eff}}
\nc{\kk}{\vek{k}}
\nc{\pp}{{\rm p}}
\nc{\ga}{g_{a\gamma}}
\nc{\vv}{\\}
\nc{\eee}{{\bf E}}
\nc{\bbb}{{\bf B}}
\nc{\qcd}{T_{\rm QCD}}
\nc{\G}{\rm \ G}
\def\vec#1{{\bf #1}}

\def\lae{\;^{<}_{\sim} \;} \def\gae{\; ^{>}_{\sim} \;} 

\def\ell{e^{c}LL}

\begin{document}
{\title{\vskip-2truecm{\hfill {{\small \\
	\hfill \\
	}}\vskip 1truecm}
{ 
Curvaton Potential Terms, Scale-Dependent Perturbation Spectra
 and Chaotic Initial Conditions}}
{\author{
{\sc  John McDonald$^{1}$}\\
{\sl\small Dept. of Mathematical Sciences, University of Liverpool,
Liverpool L69 3BX, England}
}
\maketitle
\begin{abstract}
\noindent

   The curvaton scenario predicts an almost scale-invariant spectrum of
 perturbations in most
 inflation models. We consider the possibility that renormalisable $\phi^{4}$ or 
Planck scale-suppressed non-renormalisable curvaton
 potential terms may result in an observable deviation from scale-invariance. We 
show that if the
 curvaton initially has a large amplitude and if the total number
 of e-foldings of inflation is less than about 300 then a running blue
 perturbation spectrum with an observable deviation from scale-invariance
 is likely. 
D-term inflation is considered as an example with
 a potentially low total number of e-foldings of inflation.
 A secondary role for the curvaton, 
in which it drives a period of chaotic inflation leading to
 D-term or other flat potential inflation 
from an initially chaotic state, is suggested. 

\end{abstract}
\vfil
\footnoterule
{\small $^1$mcdonald@amtp.liv.ac.uk}

\thispagestyle{empty}
\newpage
\setcounter{page}{1}

\section{Introduction}

                The curvaton scenario is an alternative model
 of the origin of cosmological density
 perturbations \cite{curv,curv2,curv3}. A scalar field, the curvaton,  is
 assumed to be effectively  massless during inflation. 
De Sitter fluctuations of the curvaton are transferred into 
adiabatic energy density perturbations 
when coherent oscillations of the curvaton field come to dominate the energy density
 of the Universe and subsequently decay. A number of particle physics candidates
 have been proposed and models analysed \cite{clist,snu}. In particular, it has been
 noted that supersymmetric (SUSY) D-term hybrid inflation \cite{dti} with natural 
values for the dimensionless couplings must have a curvaton in order to be consistent 
with cosmic microwave background (CMB) constraints \cite{kawa}.

                    A prediction of the massless curvaton model is that the spectrum of
 density perturbations is almost exactly scale-invariant in most inflation models
 \cite{curv3}. An observable scale-dependent curvaton perturbation
 ($|\Delta n| \gae 10^{-2}$, where $n$ is the spectral index \cite{obs}) could be
 obtained if
 the curvaton mass was close to $0.1 H$ during
 inflation \cite{curv3,snu}, but there is no strong reason to expect this. 

          Another possibility is that the curvaton could 
have renormalisable or Planck scale-suppressed
 non-renormalisable terms 
in its potential. It is therefore important to ask whether such
 terms could naturally result
 in an observable deviation from scale-invariance
 and to establish the nature of the deviation. 
We will consider this question in the following. We will show that 
an observable deviation from scale-invariance is possible if the total number of
 e-foldings of inflation is relatively small, less than around 300, and if the curvaton makes 
the transition to slow-rolling during inflation,
 which is likely if the initial curvaton amplitude is large. 

       D-term inflation 
provides an example of an inflation model which may have a small number of 
e-foldings of inflation. This is the case if the 
initial value of the inflaton field is less than the reduced Planck scale 
$M = M_{Pl}/\sqrt{8 \pi} = 2.4 \times 10^{18} \GeV$, implying a small 
number of e-foldings of inflation ($< 1000$) 
if the $U(1)_{FI}$ Fayet-Illiopoulos gauge coupling is larger than 0.2.  An initial 
value for the inflaton field less than the Planck scale is expected if 
dangerous non-renormalisable terms in the inflaton potential are
 suppressed by a discrete symmetry, leaving only high-order
 Planck-scale suppressed  terms.
Alternatively, in the case where inflation originates from chaotic initial
 conditions, the curvaton may play a secondary role, providing an initial period of
 chaotic
 inflation \cite{linci} which allows the transition from an initially chaotic
 state characterized by the 
Planck scale ('space-time foam' due to quantum fluctuations of the 
metric \cite{linci}) to D-term inflation (or any other inflation 
with a nearly flat potential) 
on a much lower energy scale.
 In this scenario the initial value of the inflaton field is naturally of 
the order of the Planck scale, even in the case where the inflaton
 potential has no non-renormalisable corrections. Thus a small total
 number of e-foldings of D-term inflation is also possible in this case.

    The paper is organised as folllows.  In  Section 2 we discuss the
 spectral index and deviation from scale-invariance 
due to a curvaton potential. In Section 3 we consider the specific case
 of the D-term inflation/curvaton scenario. In Section 4 we present our conclusions. 

\section{Spectral Index from Curvaton Potential Terms}

           In this section  we will consider a simple model consisting of a
constant inflaton potential, $V_{o}$, and a curvaton potential 
consisting of terms of the form,
\be{e1} V(\phi) = \frac{\lambda \phi^{d}}{d! M_{Pl}^{d-4}}    ~,\ee
where $\lambda$ is taken to be positive and of the order of 1, $d!$ is a symmetry factor and $M_{Pl} = 1.2 \times 10^{19} \GeV$. 
This includes a renormalisable $\phi^{4}$ interacton when $d=4$ and Planck 
scale-suppressed non-renormalisable terms, associated with the natural quantum gravity cut-off, for $d \geq 5$. 
The curvaton equation of motion is then,
\be{e2} \ddot{\phi} + 3 H \dot{\phi} = - \alpha_{d} \phi^{d-1}  \;\;\; ; \;\;\;
 \alpha_{d} = \frac{\lambda}{(d-1)! M_{Pl}^{d-4}}     ~,\ee
where the expansion rate $H$ is assumed to be approximately constant during inflation. 

       A natural possibility is that the curvaton has a
large initial amplitude.
For example, if all values of the initial curvaton 
amplitude are equally probable (at least for a small 
effective mass, $V^{''}(\phi) < H^{2}$), then on average we would expect to find 
a large initial amplitude. 
Alternatively, it has been suggested that the
 initial value of the potential energy density 
in quantum gravity is naturally of the order of
 $M_{Pl}^{4}$, corresponding to a chaotic initial state with quantum 
fluctuations of energy density $\sim M^{4}_{Pl}$ and size $\sim M_{Pl}^{-1}$ \cite{linci}.
 In this case a large initial
 value of the curvaton amplitude would be expected. 
The curvaton could then play a secondary role, driving a period
 of chaotic inflation
 with energy density initially of the order of $M_{Pl}^{4}$, which allows the 
 transition from the chaotic initial state to inflation at a much lower energy density. 

        For a large initial amplitude the curvaton evolution during inflation has two distinct phases,
 coherent oscillation and slow-rolling. 
A slow-rolling curvaton satisfies $\ddot{\phi} \ll 3 H \dot{\phi}$. In this limit 
\eq{e2} has the solution 
\be{e3} \phi^{d-2} = \frac{\phi_{o}^{d-2}}{\left(1
 + \frac{3 \left(d-2\right)}{\left(d-1\right)} 
\left(\frac{\phi_{o}}{\phi_{*}}\right)^{d-2} \Delta N\right)   }  ~,\ee
where 
\be{e4} \phi_{*}^{d-2} = \frac{9 H^{2}}{\alpha_{d}\left(d-1\right) }   ~\ee
and $\Delta N$ is the number of e-foldings of inflation that have passed since $\phi$ was equal to $\phi_{o}$. 
$\ddot{\phi}$ can be obtained in terms of $\dot{\phi}$ by taking the time derivative
 of \eq{e2} in the slow-roll limit,
\be{e4a}     \ddot{\phi} = -\frac{\alpha_{d} \left(d-1\right)}{3H} \phi^{d-2}
 \dot{\phi}   ~.\ee
 The condition for a slow-rolling curvaton, $\ddot{\phi} \ll 3 H \dot{\phi}$, then
 becomes $\phi^{d-2} \ll \phi_{*}^{d-2}$, such $\phi_{*}$ is approximately the
 value of $\phi$ at which curvaton slow-rolling begins.

     The spectral index of a slowly rolling curvaton is given by 
 \cite{curv3}
\be{e5} n = 1 - 2 \epsilon + \frac{2}{3} \frac{V^{''}\left(\phi\right)}{H^{2}}     
 ~,\ee
where $\epsilon = - \dot{H}/H^{2}$ and all terms are evaluated when 
the scale of present wavenumber $k$ leaves the horizon. The $\epsilon$ term is 
negligible in most inflation models \cite{lr}. The second term is due to the 
evolution of the curvaton during inflation. 
Applying \eq{e5} to \eq{e3} (with $\epsilon = 0$) gives
\be{e6} \Delta n \equiv n-1 =  \frac{6 \gamma^{d-2}}{
\left(1 + \frac{3 \left(d-2\right)}{\left(d-1\right)}\gamma^{d-2} \Delta N \right) } 
 \;\;\; ; \;\; \gamma = \frac{\phi_{o}}{\phi_{*}}  ~.\ee
Thus a curvaton slowly rolling in the potential of \eq{e1} produces
 a blue perturbation spectrum ($n > 1$). 
In addition, the running of the spectral index with present wavenumber $k$
 is given by
\be{e6a} \frac{d n}{d \ln k} = \frac{d n}{d \Delta N}
 = - \frac{\left(d-2\right)}{2\left(d-1\right)}\Delta n^{2}  ~,\ee
where $N$ is the number of e-foldings until the end of inflation and we have used $d \Delta N/dk = -d N /dk  = 1/k$. 
(The present wavenumber $k$ is proportional to $e^{-N}$, where $N$ is 
the number of e-foldings before the end of inflation at which the perturbation crosses the horizon. 
Therefore $dN/dk = -1/k$.) 

        We next consider the number of e-foldings of inflation over
 which a slow-rolling curvaton can produce a potentially observable deviation from
 scale-invariance. We will consider as examples the case $d =4$, corresponding to a 
renormalisable $\phi^{4}$ interaction, and $d = 6$ for the non-renormalisable curvaton
 potential. (The $d = 6$ case is the natural lowest order term in
 supersymmetry, corresponding to a non-renormalisable superpotential term $\propto 
\phi^{4}$.) We will consider the slow-rolling curvaton approximation to
 be reasonably well satisfied once $3 H \dot{\phi} \geq 4 \ddot{\phi}$, which
 requires that $\gamma^{d-2} \leq 0.25$. Thus the initial value of $\phi$ we will
 consider 
corresponds to $\phi_{o}$ such that $\gamma^{d-2} = 0.25$. 
For the case $d=4$
this implies that $\gamma = 0.5$ (i.e. $\phi_{o} 
= 0.5 \phi_{*}$) \footnote{For $d=4$ we have $\phi_{*} \approx H$. Therefore 
a large curvaton amplitude at the end of inflation is 
possible even with a renormalisable $\phi^{4}$ potential.}. Then 
\be{e7} \Delta n =  \frac{1.5}{\left(1 + 0.5 \Delta N \right)}   ~.\ee
Requiring that $\Delta n \geq 10^{-2}$ then requires that $\Delta N \leq 298$. 
Thus for $d=4$ an
 observable deviation from a scale-invariant perturbation spectrum is
 obtained during the first 298
 e-foldings of inflation following the onset of curvaton slow-rolling. 
For the case $d=6$ we find $\gamma = 0.71$ and 
\be{e9} \Delta n = \frac{1.5}{\left(1 + 0.6 \Delta N \right)}   ~.\ee
Requiring $\Delta n \geq 10^{-2}$ then requires that $\Delta N \leq 248$. 
 
          From this we see that if the length scales of
 cosmological interest (those corresponding to 
CMB temperature fluctuations and large-scale structure) leave the horizon less
 than about 250 e-foldings after the onset of curvaton slow-rolling,
 and if the curvaton makes the transition to slow-rolling 
 during inflation, then we would
 expect to be able to observe a running blue spectrum
 of density perturbations, assuming a 
leading order curvaton potential term with $d \leq 6$. Any positive
 value of $\Delta n$ can be accomodated, depending on how many e-foldings of
 curvaton slow-rolling have occured when the scales of cosmological interest exit the
 horizon. 

      In the simple model we are considering here inflation begins once the curvaton
 energy density becomes less than $V_{o}$.
 We can then estimate the number of e-foldings from the onset of inflation 
until the onset of curvaton slow-rolling. For $\phi > \phi_{*}$ the curvaton will have
 a large effective mass, $V^{''}(\phi) \gg H^{2}$. Therefore we expect the curvaton to
undergo damped coherent oscillations in a $\phi^{d}$ potential. 
We expect the transition from coherent oscillations to slow-rolling to occur rapidly as 
a function of $\phi$ in a $\phi^{d}$ ($d \geq 4$)
 potential. Therefore we will consider 
the curvaton to be coherently oscillating for $\phi \gae \phi_{*}$ and slow-rolling for 
$\phi \lae \phi_{*}$. 
The amplitude of coherent oscillations in a $\phi^{d}$ potential evolve with
 scale factor $a$ as 
$\phi \propto a^{-\frac{6}{d+2}}$ \cite{turner}, thus the energy density evolves as 
$V(\phi)  \propto  a^{-\frac{6 d}{d+2}}$. Therefore the 
number of e-foldings from the beginning of inflation ($V(\phi) = V_{o}$) 
until the onset of curvaton slow-rolling, $N_{S}$, is 
\be{e10}   N_{S} = \frac{\left(d +2\right)}{6d} \ln \left(
 \frac{V_{o}}{V\left(\phi_{*}\right)}\right)  ~,\ee
where the ratio of the energy density of the inflaton to the energy
 density of the curvaton at $\phi_{*}$ is given by 
\be{e11} \frac{V_{o}}{V\left(\phi_{*}\right)} \approx
 \left(\frac{M_{Pl}^{4}}{V_{o}}\right)^{\frac{2}{d-2}}     ~.\ee
(This also shows that at the onset of curvaton slow-rolling the energy density is typically
 inflaton dominated.)  Therefore, using as an example $V_{o}^{1/4} = 10^{15}
 \GeV$, we find for $d=4$ ($d=6$) that $N_{S} \approx 9.4$ ($N_{S} \approx 4.2 $). 

      Thus the total number of e-foldings from the onset of inflation during
 which an observable deviation from scale-invariance is obtained is $N_{S} +
 \Delta N \approx 310$ for $d=4$ and $\approx 250$ for $d=6$. Therefore, since the 
scales of
 cosmological interest leave the horizon at around 50 e-foldings before the end of
 inflation, if the total number of
 e-foldings of inflation is less than about 360 ($d=4$)
 or 300 ($d=6$), and if the initial
 value of the curvaton is large such that the transition
 to curvaton slow-rolling occurs during inflation, then we
would expect to find a running blue perturbation
 spectrum with an observable deviation from scale-invariance.

\section{D-term Inflation/Curvaton Scenario} 

            A possible 'application' of the curvaton scenario is to 
 SUSY hybrid inflation models.  
Hybrid inflation models \cite{hi} are a favoured class of inflation model, 
due to their ability to account both for a sufficiently flat potential during inflation 
and a sufficiently large inflaton mass for reheating after inflation, 
without requiring small couplings. In the context
 of SUSY, 
a particularly interesting class of hybrid inflation model is 
D-term inflation \cite{dti}. 
This is because D-term inflation naturally evades the $\eta$-problem of supergravity (SUGRA)
inflation models \cite{eta}
 i.e. the generation of order $H$ SUSY-breaking mass terms due to 
non-zero F-terms during inflation. Thus D-term inflation models are favoured, 
at least as low-energy effective theories. 

       However, the high precision observations
 of CMB temperature fluctuations made by the Wilkinson Microwave
 Anisotropy Probe (WMAP) \cite{wmap} have introduced a difficulty for D-term 
inflation models. The contribution of 
$U(1)_{FI}$ cosmic strings, formed at the end of 
inflation, to the CMB 
is too large unless the superpotential coupling satisfies 
$\lambda \lae 10^{-4}$ \cite{kawa}. This is 
an unattractive possibility if the 
motivation for hybrid inflation models is that
inflation can proceed with natural values 
of the dimensionless coupling constants
of the order of 1. 

      The cosmic string problem and constraint on $\lambda$ arises if the 
energy density perturbations responsible for structure formation are due to 
conventional inflaton quantum fluctuations. However, if the energy density 
perturbations were generated by a curvaton, it might be 
possible to reduce the energy density during inflation such that 
the cosmic string contribution to the CMB is acceptably small \cite{kawa}. 

         We first review the relevant features of D-term inflation \cite{dti}.
For inflaton field $s$ large compared with $s_{c}$ (where $s_{c}$ is 
the critical value of $s$ at which
the $U(1)_{FI}$ symmetry breaking 
transition occurs), the 1-loop effective potential of D-term inflation is given 
by $V = V_{o} + \Delta V$ ($ \Delta V \ll V_{o}$), where \cite{dti} 
\be{e12}  
V_{o} = \frac{g^{2} \xi^{4}}{2}  
 \;\; ; \;\;\; 
\Delta V = \frac{g^{4} \xi^{4}}{32 \pi^{2}}
 \ln\left(\frac{s^{2}}{\Lambda^{2}}\right) 
  ~,\ee
and where $\Lambda$ is a renormalisation scale, 
$g$ is the $U(1)_{FI}$ gauge coupling and 
$\xi$ is the Fayet-Illiopoulos term.   
During slow-rolling the inflaton evolves according to 
\be{e13} 3 H \dot{s} \approx-\frac{dV}{ds} =     
-\frac{g^{4} \xi^{4}}{16 \pi^{2} s} ~.\ee
This has the solution
\be{e14} s^{2} = s_{c}^{2}  + \frac{g^{4} \xi^{4} N}{24 \pi^{2} H^{2}}    
~,\ee
where $s$ is the value of the curvaton at $N$ e-foldings before the end of 
 inflation. For $s^{2} \gg s_{c}^{2}$ the
 number of e-foldings until the end of inflation is related to $s$ by
\be{e15} s^{2} = \frac{g^{2}M^{2}N}{4 \pi^{2}}    ~,\ee
where we have used $V \approx V_{o}$ ($\Delta V \ll V_{o}$) in $H$. 
Thus $s \propto \sqrt{N}$. 

     In the absence of curvaton evolution during inflation, the spectral index is given by 
\be{e16}   n - 1 = 1- 2 \epsilon \equiv 1 - \frac{1}{V}
 \frac{dV}{d N} = 1 - \frac{g^{2}}{16 \pi^{2} N}    ~.\ee
Thus with $N \approx 50$, corresponding to scales of cosmological interest, we find
$ \Delta n = n-1 =  -1.3 \times 10^{-4} g^{2}$.
Therefore the deviation from scale-invariance is unobservably small in the absence of
 curvaton evolution. 

                 For natural values of the gauge coupling $g$, the value of inflaton
 at $N \approx 50$ is close to $M$. This leads to a problem associated with
 non-renormalisable corections to the superpotential of the form 
$S^{m}/M^{m-3}$ and $U(1)_{FI}$ gauge superfield 
terms of the form 
$S^{k}W^{\alpha}W_{\alpha}/M^{k}$ \cite{kmr}. These would result in an unacceptable deviation 
of the potential from flatness at $N \approx 50$ (so preventing sufficient inflation) 
unless $m > 9$ and $k > 6$ \cite{kmr}. The possible solutions of this problem are to 
impose either an R-symmetry or a discrete symmetry on the superpotential to eliminate 
the dangerous terms \cite{kmr}. Discrete gauge symmetries are 
preferred if quantum gravity effects violate global symmetries such as R-symmetry 
\cite{kmr}. 

          A high-order discrete symmetry would eliminate 
the dangerous terms up to a large value of $m$. In this case the inflaton
 potential 
 would rapidly increase as $s$ approaches $M$. Thus we would expect that inflation starts at
 $s < M$ in the case of a discrete symmetry. In this case the total number of e-foldings of
 inflation is less than $ 4 \pi^{2}/g^{2}$. 
For $g > 0.2$ the total number of e-foldings is less than 1000 \footnote{A concern
 with high-order discrete symmetries is that 
they may suppress the reheating temperature by suppressing couplings of the inflaton to 
the Standard Model sector fields. This would make it 
difficult for the curvaton to dominate the 
energy density before it decays.}. 

            An R-symmetry would, in most cases, eliminate all 
non-renormalisable terms from the inflaton potential.
 This would result in a nearly flat inflaton
potential for all values of the inflaton. In this case the question of the initial 
value of the inflaton field is related to the dynamics of the onset of inflation. 
One possibility is that the Universe begins in a chaotic state, characterised by 
quantum fluctuations of energy density of the order of $M^{4}$ and
 length scales $M^{-1}$ \cite{linci} \footnote{We follow \cite{kyy2} 
and assume that the scale 
of chaotic fluctuations in SUGRA is set by the reduced Planck scale.}. 
The energy density will consist of 
gradient, kinetic and potential terms. For inflation to begin, it is essential that one
 of the fluctuations can become dominated by the potential term. This 
requires that inflation can occur with a potential  
energy density of the order of $M^{4}$, as in chaotic inflation \cite{linci}.
 D-term inflation with a flat potential has an
 energy density characterised by the scale $\xi \lae 10^{16} \GeV \ll M$.
 Therefore, in order to enter D-term inflation (or any other inflation 
with a nearly flat potential), an initial period of
 chaotic inflation due to another field is essential.  The curvaton field with a potential
 could play exactly this role. If the curvaton starts with a large energy density then we
 may have an initial period of curvaton-driven chaotic inflation
 followed by D-term inflation. In
 this case we also expect the transition to curvaton
 slow-rolling to occur during D-term
 inflation. The initial energy density of the inflaton
 will be due gradient terms characterised 
by the length scale $M^{-1}$, $\rho \sim M^{2}s^{2}$, assuming that the 
initial value of the inflaton field, $s$, is due to a chaotic fluctuation of 
length scale $M^{-1}$ which subsequently inflates. Since the initial gradient energy 
density is expected to be of the order of $M^{4}$, we expect that
 $s \sim M$. Therefore, with chaotic initial conditions, it is plausible
 that a small number of e-foldings of D-term inflation will occur, depending 
 on the random initial fluctuation which inflates become to the observed Universe 
\footnote{In order to drive chaotic inflation, it is necessary for the curvaton to take 
a value initially larger than $M$, which requires a non-trivial structure for
 the SUGRA K\"ahler potential \cite{kyy,kyy2}. In our case the initial 
value need not be too large, since we only require a small 
number of e-foldings of chaotic 
inflation in order to make the transition from the chaotic initial state to 
potential-driven evolution.}. 

   Thus D-term inflation with a discrete symmetry solution of the Planck-scale inflaton 
problem naturally results in a small total number of e-foldings of inflation. 
In addition, with chaotic initial conditions it is also plausible that a small number of
 e-foldings of D-term inflation will occur. 
Therefore
 an observable deviation of the curvaton perturbation spectrum from scale-invariance 
is possible in these cases.     

\section{Conclusions}

               We have shown that a curvaton scenario 
with a natural curvaton potential,
 large initial curvaton amplitude and less than 
around 300 total e-foldings of inflation can result in a blue perturbation spectrum 
with an observable deviation from scale-invariance and running spectral index. 
   
           Of the two conditions required to obtain an observable deviation
 from scale-invariance, the requirement of
 a large initial curvaton amplitude (such that the curvaton makes
 the transition to slow-rolling during inflation) is not manifestly unnatural. Indeed, it may 
 be the most natural initial condition to consider if the Universe evolves from a
 chaotic initial state. 

            The question is then whether it
 is natural to have a total number of e-foldings of inflation not much larger than 300. 
In general, there is no obvious reason to expect 
the total number of e-foldings to be as low as 300. 
However, in some models, such as D-term inflation with a discrete symmetry 
suppression
 of dangerous non-renormalisable inflaton potential terms, a low number of e-foldings
 ($\lae 1000$ for $g \gae 0.2$) is likely. 
Alternatively, if the Universe starts in an initial 
chaotic state characterised by energy density fluctuations $\sim M^{4}$ on length
 scales $M^{-1}$, then the curvaton may play a secondary role by providing an initial 
period of chaotic inflation which allows the transition from the initial chaotic state to 
D-term inflation  
at a much lower energy scale. In this case we expect the initial 
value
 of inflaton field to be of the order of $M$, which can plausibly 
result in a low total number of e-foldings of D-term 
inflation. It would be interesting to develop a complete D-term inflation/curvaton 
scenario, compatible with
 chaotic initial conditions as well as CMB constraints.  

    The idea that the curvaton may provide an initial period of chaotic inflation 
is more general than the specific application to D-term inflation, and may be 
applied to other inflation models based on a nearly flat potential. 

      The simplest interpretation of the
 results presented here is that they support the idea that the curvaton is likely to 
produce an effectively 
scale-invariant perturbation spectrum as far as observations are concerned, given that
 most inflation models will have a total number of e-foldings much larger than 300.
 However,
 if a running blue perturbation spectrum is observed, it could be interpreted as
 the effect of a curvaton potential combined with
 a relatively small number of e-foldings of inflation, requiring a particular 
inflation model such as D-term inflation with a high-order discrete symmetry or 
chaotic initial conditions and an initial period of curvaton-driven chaotic inflation.

\end{document}